\documentclass[aps,twocolumn,reprint,amsmath,amssymb,showpacs,superscriptaddress]{revtex4-1}
\usepackage{multirow}
\usepackage{graphicx,epsfig}
\usepackage{wasysym}
\usepackage{color,soul} 
\usepackage[colorlinks=true,citecolor=blue,linkcolor=red,linktocpage=true,pagebackref=false]{hyperref}

\def \beq {\begin{equation}}
\def \edq {\end{equation}}
\def \ba {\begin{eqnarray}}
\def \ea {\end{eqnarray}}
\def \bes {\begin{subequations}}
\def \eds {\end{subequations}}
\def \beqn {\begin{equation*}}
\def \edqn {\end{equation*}}

\def \dag {\dagger}

\def \veps {\varepsilon}

\def \calh {{\cal{H}}}

\def \calg {{\cal{G}}}

\begin{document}
\title{Dynamics of energy transport and entropy production in ac-driven quantum electron systems}
\author{Mar\'{\i}a Florencia Ludovico}
\affiliation{Departamento de F\'{\i}sica, FCEyN, Universidad de Buenos Aires and IFIBA, Pabell\'on I, Ciudad Universitaria, 1428 CABA Argentina}
\affiliation{International Center for Advanced Studies, UNSAM, Campus Miguelete, 25 de Mayo y Francia, 1650 Buenos Aires, Argentina} 
\author{Michael Moskalets}
\affiliation{Department of Metal and Semiconductor Physics,
NTU "Kharkiv Polytechnic Institute", 61002 Kharkiv, Ukraine}
\author{David S\'anchez}
\affiliation{Instituto de F\'{\i}sica Interdisciplinar y Sistemas Complejos
IFISC (UIB-CSIC), E-07122 Palma de Mallorca, Spain}
\author{Liliana Arrachea}
\affiliation{Departamento de F\'{\i}sica, FCEyN, Universidad de Buenos Aires and IFIBA, Pabell\'on I, Ciudad Universitaria, 1428 CABA Argentina}
\affiliation{International Center for Advanced Studies, UNSAM, Campus Miguelete, 25 de Mayo y Francia, 1650 Buenos Aires, Argentina}

\begin{abstract}
We analyze the time-resolved energy transport and the entropy production in ac-driven quantum coherent electron systems coupled to multiple reservoirs at finite temperature. 
At slow driving we formulate the first and second laws of thermodynamics valid at each instant of time.  
We identify heat fluxes flowing through the different pieces of the device and emphasize the importance of the energy stored in the contact and central regions for the second law of thermodynamics to be instantaneously satisfied. 
In addition, we discuss conservative and dissipative contributions to the heat flux and to the entropy production as a function of time. We illustrate these ideas with a simple model corresponding to a driven level coupled to two reservoirs with different chemical potentials.
\end{abstract}


\pacs{73.23.-b, 72.10.Bg, 73.63.Kv, 44.10.+i}
\maketitle

\section{Introduction} 
The understanding of the energy transfer in non-equilibrium open quantum systems  is a fundamental problem in physics. The separation of energy  in heat  and useful work  and dissipation is the key for a thermodynamical description.  In quantum systems under ac driving, the identification of these different components of energy is a non-trivial task which is paramount to cold atoms \cite{cold}, 
nanomechanical  \cite{nanomec1,nanomec2},
nanoscale optoelectronical \cite{Michelini:2016ju}, 
and mesoscopic electron physics \cite{elec1,elec2,wu09,elec3,elec4,Carrega:2015dh,Chen:2015hu, camp:2015,thermolud,Proesmans:2016tq,Dare:2016jk,Bruch:2015ux}. 
 Typically, the central piece of these systems contains a small number of particles and are driven out of equilibrium, which renders a usual thermodynamical description unreliable. However they are in contact to one or more macroscopic reservoirs with well defined thermodynamical intensive parameters. 

In the recent years, the name ''quantum thermodynamics''  has been coined to identify  the area of Physics devoted to the study of this type of systems, which is an intersection of solid state and statistical physics. The foundations of this area were in part developed after the proposal of the Jarzynski's   equality \cite{jar}
and
Crook's theorem \cite{crook} and a subsequent number of fluctuations relations \cite{fluc1,fluc2,sai08,for08,san09,fluc3,fluc4,nak10,fluc5,lop12,wan13,uts14}. Recently, linear response proposals in close relation to thermodynamics have been formulated for open quantum systems and quasi-classical systems under periodic driving \cite{thermolud, thermo1,thermo2,thermo3}.  The proper definition of the heat exchange between a quantum driven system and its 
macroscopic environment
 has been recently addressed
in the context of few-level or spin systems in contact to phononic baths \cite{heat1, heat2, heat3}  and in systems of coupled quantum harmonic oscillators \cite{heat4,heat5,heat6}.

The first law of thermodynamics, being basically the conservation of the energy, is equally valid for  non-equilibrium and equilibrium phenomena.  
We have recently considered a model containing the minimal ingredients to address the problem of time-resolved heat transport~\cite{us}. It consists of a localized level  under ac driving coupled to a single electron reservoir.  
We have focused on slow driving and zero temperature. 
By slow we mean a regime where the typical dwell time for the electrons inside the driven structure is much smaller than the driving period.
Even in such a simple setup a nontrivial effect manifests itself when the heat flow is analyzed as a function of time. Namely,   the coupling region between the different parts of the system  behaves like an {\em energy reactance}. 
In this way, the coupling not only provides a necessary mechanism for particle and energy exchange but also contributes to the energy balance.
This contribution is of ac nature. It allows for a temporary energy storage which vanishes when averaged over time.

\begin{figure}[h!]
  \includegraphics[scale=0.35]{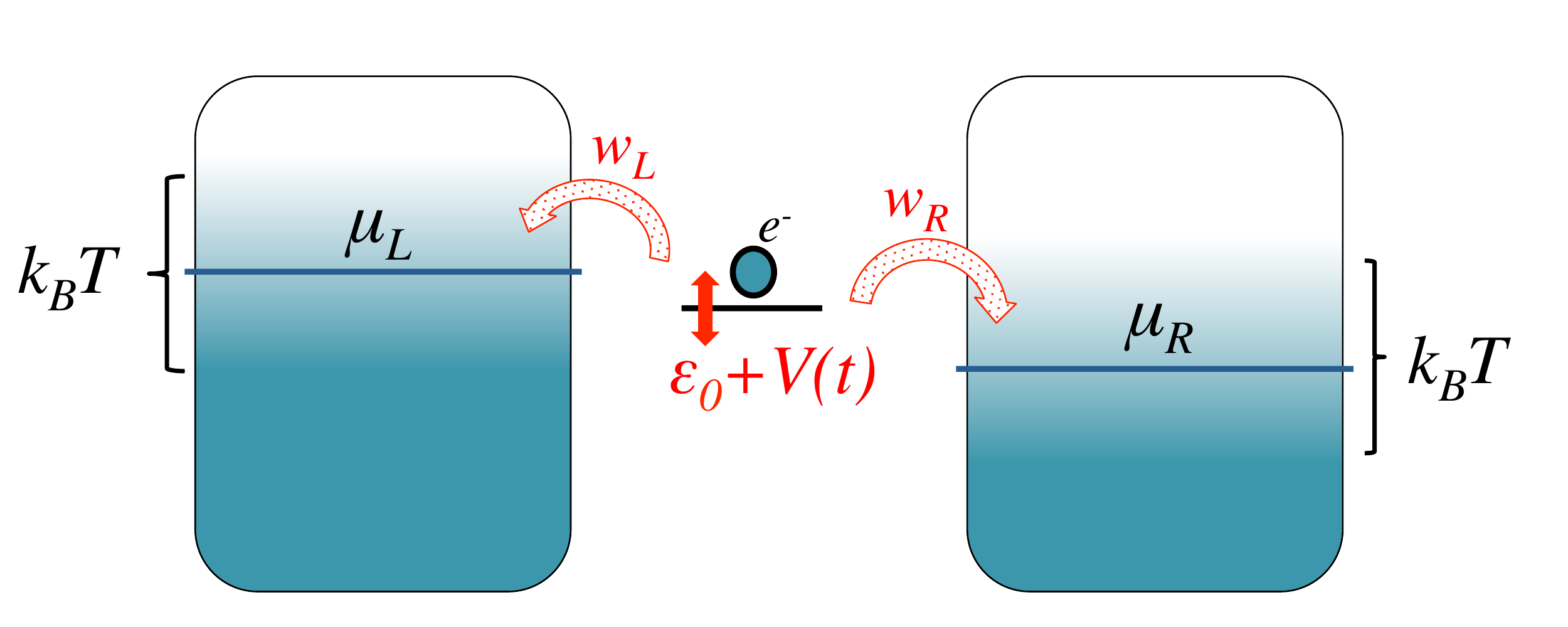}
  \caption{A single electronic level is coupled to two reservoirs (fermionic baths) kept at the same temperature $T$. The chemical potentials of the left and right reservoirs are $\mu_L=\mu$ and $\mu_R=\mu-\delta\mu$, respectively. The electronic level slowly evolves in time with a periodic parameter $V(t)$, and hence after a completed period the central part of the systems returns to its initial state.}  
  \label{fig1}
\end{figure}

Our goal now is to analyze the time-resolved energy redistribution and entropy production in ac-driven quantum coherent electron systems coupled to multiple reservoirs and finite
temperature. 
 We show that the definition of the heat current flowing into the reservoirs presented in Ref.~\cite{us} is also suitable for multi-terminal devices. More interestingly, we study the behavior of the different components of the heat. 
We identify conservative and dissipative contributions to a heat flux and to the entropy production as a function of time. 
We illustrate these ideas with a simple system that consists of a slowly driven resonant level coupled to two electron reservoirs at a finite temperature and with an applied bias voltage, see Fig.~\ref{fig1}. 

The paper is organized as follows. We present the model in Sec.~II. A thermodynamic approach to the case of slight departures from equilibrium is presented in Sec.~III. Section IV contains the definition of the time-dependent energy fluxes, conservation laws and  the different contributions to the entropy production. In Sec.~V we focus our analysis on the slow driving regime. These ideas are then illustrated in Sec.~VI, where the example of a single driven level coupled to reservoirs is analyzed in detail. Section VII is devoted to the summary and conclusions.

\section{Model}
 We consider a finite quantum system, as for example a single quantum dot or an array of quantum dots, which is driven out of equilibrium by time-periodic adiabatic power sources and in contact to several fermionic baths. 
Then,
the Hamiltonian of the full system can be separated into three contributions,
\beq\label{eq_H}
\calh(t) = \calh_{\rm res} +\calh_S(t)+ \calh_{\rm cont} .
\edq 

The Hamiltonian representing the reservoirs (fermionic baths) is $\calh_{\rm res}= \sum_{\alpha} \calh_{\alpha} $ with 
$ \calh_{\alpha}= \sum_{ k_{\alpha}}\veps_{k_{\alpha}}c_{k_{\alpha}}^{\dag}c_{k_{\alpha}}$,
where $\veps_{k_{\alpha}}$ is the energy dispersion relation and $c_{k_{\alpha}}^{\dag}$ ($c_{k_{\alpha}}$)
creates (destroys) an electron with continuous index (wavenumber) $k_{\alpha}$.
Each of these reservoirs is at local equilibrium
with a well defined temperature $T$ and chemical potential $\mu_{\alpha}$.
The Hamiltonian $\calh_S( {\bf V}(t) )$ describes the central piece of the setup, where electrons are confined and the driving is applied. For generality, the form of  $\calh_S( {\bf V}(t) )$ remains unspecified.
The time dependence is introduced via a set of parameters 
${\bf V}(t) ={\bf V}(t+\tau)= (V_1(t), \ldots, V_M(t))$ 
which characterize the sources of the ac driving, with $\tau$ being the driving period. Finally, the term 
$\calh_{\rm cont} =  \sum_{\alpha} \calh_{c \alpha} $ with
\beq\label{eq_contacts}
 \calh_{c \alpha}= \sum_{k_{\alpha}, l_{\alpha}}\left( w_{k_{\alpha},l_{\alpha}} c^{\dagger}_{k_{\alpha}} d_{l_{\alpha}} + H. c \right),
 \edq
  describes the tunneling hybridization between the electrons at the reservoirs and the central system. This tunneling takes place in a contact region that separates the reservoirs and the central piece. In Eq.~\eqref{eq_contacts} the fermionic operators $d_{l_{\alpha}}$ and  $d^{\dagger}_{l_{\alpha}}$ are associated to the degrees of freedom of the central system. 
  In what follows, we present a general reasoning, which is valid for any  $\calh_S({\bf V}(t))$, even when the central piece contains many-body interactions.

\section{Thermodynamic approach} \label{thermo}
The aim of this section is to present a treatment similar to the one presented in Ref.~\onlinecite{balian} in order to identify heat and work and express the first and the second laws of thermodynamics
in a process involving small deviations from equilibrium due to slow variations of the time-dependent parameters $\delta {\bf V}$ entering ${\cal H}$.
 
\subsection{Entropy and the first law}\label{b}
\subsubsection{Reservoirs with equal  chemical potentials}

Let us begin discussing the case $\mu_{\alpha}=\mu$ and $T_{\alpha}=T$. 
For an equilibrium system a description based on the grand canonical ensemble such that $\rho= e^{-\beta( {\cal H}-\mu {\cal N})}/Z$, 
with $Z= \mbox{Tr}\left[  e^{-\beta( {\cal H}-\mu {\cal N})} \right]$ the partition function, ${\cal N}$ the particle number and $\beta=1/(k_BT)$ with $k_B$ being the Boltzmann's constant, has an associated von Neumann entropy
\beq \label{ent0}
S=-k_B\mbox{Tr}\left[  \rho \ln \rho \right] .
\edq

We now consider entropy variations that arise from small but explicit changes 
in the Hamiltonian $\delta {\cal H}= (\partial {\cal H}/\partial {\bf V}) \delta {\bf V} $ due to the variation in time of the parameters ${\bf V}$. Such variations take place within a short time interval $\delta t$ and assume that the net change $\delta {\bf V}={\bf V}(t+\delta t)-
{\bf V}(t)$ is small
compared to the typical energies (e.g., typical level spacing) of the system. The consequent change in the probability distribution is quantified by $\delta \rho= \rho(t+\delta t) - \rho(t)$ while the change in the entropy is
\beq \label{ent}
\delta S= \frac{1}{T} \mbox{Tr}\left[  \delta \rho( {\cal H} -\mu {\cal N}) \right] -  \frac{1}{T}
{\bf F} \cdot \delta {\bf V} ,
\edq
where we have defined the force 
\beq
{\bf F} =  - \mbox{Tr}\left[  \rho \frac{\partial {\cal H}}{\partial {\bf V}} \right],
\edq
and also used that $\mbox{Tr}\left[ \delta \rho \right]=0$, which is a consequence of the normalization of the probability distribution.
Here, the trace is evaluated with respect to the eigenvalues $\vert m(t)\rangle$ of ${\cal H}$ at the time $t$ with eigenenergies $E_m$. As in Ref. \onlinecite{balian} 
we have introduced the `adiabatic'' approximation, in which
$\vert\langle m^{\prime} \vert \partial {\cal H}/\partial t \vert m \rangle \vert \ll (E_m- E_{m^{\prime}})^2/\hbar $ and $\delta E_m = \langle m \vert \delta {\cal H} \vert m \rangle$.

 In Eq. (\ref{ent}), we can identify the term 
 \beq
 \delta U =  \mbox{Tr}\left[  \delta \rho {\cal H} \right]= \sum_{\alpha}\left[ \delta U_{\alpha} + \delta U_{c \alpha} \right] + \delta {U}_{S},
 \edq
 with the variation of the internal energy stored in the full system, including the variation in the central system $\delta {U}_{S}$, plus the reservoirs $\delta U_{\alpha}$ and the contact regions $ \delta U_{c \alpha} $. 

The different contributions to the variation of the internal energy are
\beq
 \delta U_{\nu} =   \mbox{Tr}\left[  \delta \rho {\cal H}_{\nu} \right],\; \nu=\alpha, c\alpha, S.
 \edq
Similarly, the variations of the number of particles stored in the different parts of the setup are
\beq
\delta N_{\nu} =  \mbox{Tr}\left[  \delta \rho {\cal  N}_{\nu} \right],\; \nu=\alpha, S
\edq
and the total change reads
\beq
\delta N = \sum_{\alpha} \delta N_{\alpha} + \delta N_S.
\edq
Crucially, the contact regions described by Eq.~\eqref{eq_contacts} have an associated energy term that
will contribute to the energy flux. In contrast, the reservoir and the system have {\it both} energy and particle
terms. Hence, the study of energy dynamics is fundamentally distinct from its particle counterpart
because one must consider the intermediate regions that partition the central system from the reservoirs. 

The second term of Eq.~(\ref{ent}) is the work done by the ac forces
 \beq
\delta W_{\rm ac}= {\bf F} \cdot \delta {\bf V}.
\edq
Hence
 \beq \label{eq_FTL}
 T \delta S = \delta U - \mu \delta N - \delta W_{\rm ac} = \delta Q_{\rm tot},
 \edq
 leads us to identify the total heat as
\beq \label{delqt}
\delta Q_{\rm tot} = \sum_{\alpha}\left[ \delta U_{\alpha} + \delta U_{c \alpha} - \mu \delta N_{\alpha}  \right] + \delta {U}_{S} - \delta W_{\rm ac}- \mu \delta N_S.
\edq

Equation~(\ref{eq_FTL}) is a statement of the first law of thermodynamics. Now, in our full system the total internal energy remains constant in a process where the central system changes due to a change of the parameters $\delta {\bf V}$. In such a process, there is an 
exchange of internal energy between the different pieces of the system but the total internal energy is conserved. The same remark applies to the total number of particles. Hence,
\begin{subequations}
\ba \label{cons}
\delta U&= & \sum_{\alpha} \left[  \delta U_{\alpha}+ \delta U_{c, \alpha} \right] + \delta U_S= 0, \label{cons1} \\
\delta N&= & \sum_{\alpha} \delta N_{\alpha} + \delta N_S=0.\label{cons2}
\ea
\end{subequations}
Therefore, we have
\beq
\delta Q_{\rm tot}= - \delta W_{\rm ac},
\edq
indicating that all the work developed by the external ac sources is transformed into heat that is absorbed by the full system containing reservoirs, central part and contacts. Notice, however, that the assumption of a constant temperature for the reservoirs implicitly assumes that they are indeed in contact to an extra bath, where the heat is finally released.

\subsubsection{Reservoirs with different chemical potentials}
We turn to consider the situation where the temperature is kept constant
but there is now a small bias in the chemical potentials of the reservoirs so that $\mu_{\alpha}= \mu + \delta \mu_{\alpha}$. 

In this
more general situation we can proceed as in the previous section to derive the contribution to the total heat generation due to the ac forces
Eq. (\ref{delqt}). However, in the present case we must also consider an additional change in the entropy $\delta S_{\rm el}$ due to the electrical work 
\beq
\delta W_{\rm el}=
\sum_{\alpha} \delta \mu_{\alpha} \delta N_{\alpha},
\edq
which is done by external batteries to maintain the bias in the electrochemical potential $\delta \mu_{\alpha}$ at the reservoirs. As before, we assume that this change is small enough to imply a slight departure from equilibrium. 
In the absence of ac forces, we have
$T\delta S=T \delta S_{\rm el}= -\delta W_{\rm el}$. Therefore when we consider the effect of the
ac voltage along with the effect of a small change in the electrochemical potentials at the reservoirs, we have to add the term $T \delta S_{\rm el}$ to Eq. (\ref{delqt}). 
 This leads to the definition of the total heat as
\ba \label{delqtg}
\delta Q_{\rm tot} & = &  \sum_{\alpha}\left[ \delta U_{\alpha} + \delta U_{c \alpha} - \mu_{\alpha} \delta N_{\alpha}  \right] + \delta {U}_{S}  - \mu \delta N_S \nonumber \\
& & - \delta W_{\rm ac}.
\ea
Assuming that the bias generates a redistribution of the particles within the setup preserving the total number of particles of the full system, Eqs.~(\ref{cons}) hold. Then,
we have
\beq \label{delqtg1}
\delta Q_{\rm tot} = - \delta W_{\rm ac}- \delta W_{\rm el}.
\edq

\subsection{Reversible and irreversible processes. The second law} \label{c}
Quite generally, all the forces developing some work can be classified as conservative and dissipative. This applies to those generating the ac driving, identified with
 $\delta W_{\rm ac}$,
as well as those corresponding to the electromotive forces by dc batteries, identified with 
$\delta W_{\rm el}$.  Hence, the heat  contains a reversible component associated to the work developed by the conservative forces, as well as  a dissipative component associated to the non-conservative forces,
\beq
\delta Q_{\rm tot}= \delta Q^{\rm rev}_{\rm tot}+ \delta Q^{\rm diss}_{\rm tot}.
\edq
In a purely reversible process, which consists of a sequence of equilibrium states defined with a density $\rho^{f}$ given by ${\cal H}(t)$,
and $\dot{\bf V} \rightarrow 0$, we have
\beq
\delta S^{\rm rev}=-\frac{\delta W^{\rm cons}}{T}= \frac{\delta Q^{\rm rev}_{\rm tot}}{T}.
\edq 
Under a cycle, which begins and ends at the same equilibrium state, $\delta S^{\rm rev}=0$, while for a general change, the quantity can take any sign. There is, however, no contradiction with the second law,
since any of such processes is akin to the isothermal expansion or compression of a gas in contact to reservoirs. In fact, in the present case, we are assuming that the reservoirs remain at the same temperature under the change. As in the case of the gas, there is still some external agent other than the reservoirs defined in the system which invests an extra work in order to maintain the temperature of the reservoirs. When taking this
action also into account, the total entropy always increases or remains constant.
Similarly the change of the entropy associated to the dissipative component is 
\beq
\delta S^{\rm diss}=-\frac{\delta W^{\rm diss}}{T}= \frac{\delta Q^{\rm diss}_{\rm tot}}{T}.
\edq
This component accounts for irreversible processes and has a non-vanishing mean value when averaged over a cycle.

\section{Kinetic approach} \label{kinetic}

Our aim now is to define fluxes that determine the rate of change of the energy and of the number of particles for different parts of the system. 
We then identify the component of the energy flux corresponding to heat and the one corresponding to work.
In addition, we will discuss the possibility of identifying fluxes of heat and work corresponding to the dynamics of the energy flow through different parts of the device.
All the equations presented in this section are exact and valid for any amplitude and frequencies of the driving potentials, degree of coupling between the system and the reservoirs, and
model Hamiltonian.   In the case of reservoirs at zero temperature, the symbol $\langle É. \rangle $ denotes the exact expectation value with respect to the exact quantum mechanical state of the full system at time 
$t$. For reservoirs at finite temperature, they correspond to the exact statistical averages with suitable exact mixed states at time $t$.  A usual procedure to
evaluate those averages is to start with the system uncoupled from the reservoirs at $t=-\infty$ and to adiabatically connect the reservoirs and the central piece of the setup. 
The exact evolution of the mean values of the observables 
of interest can be done, for instance, by recourse to Keldysh non-equilibrium Green's functions \cite{ramer}. We will focus on the state for which the evolution does not depend on the details of
the switching-on protocol for the contacts. Notice that, due to the time-dependent periodic driving, this state is also periodic in time.
In this section, we will not address the particular procedure followed to carry out the evaluation of the different mean values but rather focus on the derivation of
exact  equations relating the different rates and fluxes.

\subsection{Conservation Laws}
For any driven system described by the Hamiltonian given by Eq.~(\ref{eq_H}) we can write down two fundamental  laws: {(i)} instantaneous conservation of charge and (ii) instantaneous conservation of energy. 

(i) The total charge of the system is related to the number of particles $N$ within the whole system and the corresponding change can be expressed in terms of the variations of charge in the reservoirs and the system $I_{\nu}^C(t)=e\dot{\langle {\cal N}_{\nu} \rangle }=  \frac{ie}{\hbar}\langle \left[ {\calh}, {\cal N}_{\nu} \right] \rangle $, with $\nu=\alpha,S$,
\beq \label{numbal}
e\dot{ \langle {\cal N} \rangle } =  I^C_S(t)+\sum_{\alpha} I^C_{\alpha}(t).
\edq 
$I^C_{\alpha}$ are effectively charge currents that flow into or out of the reservoirs while $ I^C_S(t)$
can be interpreted as a displacement current which is finite only in time-dependent situations, like the stationary time-periodic regime we are addressing.

Charge conservation implies that $\dot{ \langle {\cal N} \rangle }=0$ and then we obtain an instantaneous balance for the electric currents
\beq \label{numbal2}
 \sum_{\alpha} I^C_{\alpha}(t)+I^C_S(t)=0.
\edq

(ii) To analyze the equation for the dynamics of the energy exchange between the different parts of the system
 we
define the following energy fluxes
\beq \label{encur}
J^E_{\nu}(t) =  \frac{i}{\hbar} \langle \left[ {\calh}, {\calh}_{\nu} \right] \rangle ,
\edq
 with $\nu \equiv \alpha, c \alpha, S$, which are understood as energy variations corresponding to the reservoir, the contact and the central piece, respectively.  We also define the generalized force 
 \beq \label{force}
 {\bf F} = - \left\langle \frac{ \partial \calh}{\partial {\bf V}} \right\rangle.
 \edq 
Now, we can
 derive the following exact equations
 \ba
  \dot{ \langle {\cal H}_{\alpha} \rangle } &=& J^E_{\alpha}(t)   , \label{jeal} \\
 \dot{ \langle {\cal H}_{c \alpha} \rangle } &= & J^E_{c \alpha}(t) = - J^E_{\alpha}(t) + \frac{i}{\hbar} \langle \left[ {\cal H}_S, {\cal H}_{c \alpha} \right] \rangle \nonumber \\
 & & 
 +  \frac{i}{\hbar}  \sum_{\beta}  \langle \left[ { \cal H}_{c \beta}, {\cal H}_{c \alpha} \right] \rangle,  \label{jec}\\
 \dot{ \langle {\cal H}_{S} \rangle } & = & J_S^E(t)- {\bf F} \cdot \dot{\bf V}. \label{hsdot}
 \ea
 Equation~(\ref{jec}) implies
 \beq \label{intcon}
\sum_{\alpha} \left[  J^E_{\alpha}(t) + J^E_{c \alpha}(t) \right] + J^E_S(t) =0
\edq
We note that Eq.~(\ref{intcon}) is the counterpart of the first conservation equation Eq.~(\ref{cons1}) while Eq.~(\ref{numbal2}) corresponds to the second conservation equation, Eq.~(\ref{cons2}).

To evaluate the change in time for the total energy associated to the full Hamiltonian ${\calh}$ we must add the contributions of Eqs.~(\ref{jeal}), (\ref{jec}) and (\ref{hsdot}). This leads to
\beq \label{enbal}
\dot{  \langle \calh \rangle }= \sum_{\alpha} \left[  J^E_{\alpha}(t) + J^E_{c \alpha}(t) \right] + J^E_S(t) - 
{\bf F} \cdot \dot{\bf V}.
\edq
Notice that, in contrast to the charge, the energy due to a change in ${\cal H}$ is not conserved. This is because such a change corresponds to a change in internal energy of the electrons as well as the work done by the ac forces. Hence, the corresponding rate of change is equal to the power developed by the ac sources. In fact, substituting Eq.~(\ref{intcon}) into Eq.~(\ref{enbal}) we find
\beq \label{enbal2}
P_{\rm ac}(t)= - \dot{ \langle \calh \rangle } = {\bf F}\cdot \dot{\bf V}.
\edq

Interestingly, when we consider time-averaged quantities defined as 
$\overline{O}=\lim_{\tau\to \infty}\left(\int_0^{\tau}Odt\right)/\tau$, we obtain $ \overline{\dot{\langle {\cal N}_S \rangle}}=  \overline{\dot{\langle {\cal H}_S \rangle}} = \overline{\dot{\langle {\cal H}_{c \alpha} \rangle}} =0$. 
Mathematically this follows from the fact that the quantities $\langle {\cal N}_S \rangle$, $\langle {\cal H}_S \rangle$, and $\langle {\cal H}_{c \alpha} \rangle$ are bounded while $\tau \to \infty$. 
Physically this follows from the fact that charge and energy can be stored or sunk at a net rate only at the reservoirs. 

Then, the  conservation laws for the averaged quantities read
\beq \label{av1}
\sum_{\alpha} \overline{J^E_{\alpha}} =-\overline{J^E_{S}} = -\overline{P_{\rm ac}},\;\;\;\;\;\;\;\;\;\;\;\; \sum_{\alpha} \overline{I^C_{\alpha}} =0,
\edq
since
\beq \label{av2}
 \overline{J^E_{c \alpha}} = \overline{I^C_{S}}=0,
\edq
which means that there are components of the fluxes that contribute purely dynamically but do not lead to any dc contribution
in the stationary state, thereby the term \textit{reactance}.

\subsection{Defining total heat and work fluxes} \label{a}

In the case of bias voltages applied to the reservoirs $\delta \mu_{\alpha}$ through $\mu_{\alpha}= \mu + \delta \mu_{\alpha}$, the power developed by the electromotive forces in the presence of a charge flux $I^C_{\alpha}(t)$ is
\beq \label{pva}
P_{\alpha} (t)= \frac{I_{\alpha}^C(t)}{e} \delta \mu_{\alpha}.
\edq

We now turn to explore the proper definition of heat. To this end, we  consider the case where the reservoirs are at the same temperature $T$, but they have different chemical potentials.
We can perform the following operation: calculate Eq.~(\ref{enbal})$-(\mu/e)$Eq.~(\ref{numbal2}), use Eq.~(\ref{enbal2}) and collect terms conveniently to write
\ba \label{prin1}
& & \sum_{\alpha} \left[J^E_{\alpha}(t)- \mu_{\alpha} \frac{I^C_{\alpha}(t)}{e} +J^E_{c\alpha}(t) \right]+
J^E_S(t)\nonumber\\
& & \;\;\;\;\;\;\; -\mu \frac{I^C_S(t)}{e}  +P_{\rm el}(t) =0,
\ea
where 
\beq
P_{\rm el}(t)=\sum_{\alpha}P_{\alpha} (t)
\edq
is the total power developed by the electromotive forces represented by $\delta\mu_\alpha$. By comparing with Eq.~(\ref{delqtg}) we observe that we can define
 the total heat variation as
 \ba
 \dot{Q}_{\rm tot}(t) &=& \sum_{\alpha}  \left[J^E_{\alpha}(t)- \mu_{\alpha} \frac{I^C_{\alpha}(t)}{e} +J^E_{c\alpha}(t) \right]+ J^E_S(t)  \nonumber \\
 & &- P_{\rm ac}(t) -\mu \frac{I^C_S(t)}{e}.
 \ea
 Then, using Eq.~(\ref{prin1}) as well as  the conservation laws Eq.~(\ref{numbal2}) and Eq.~(\ref{intcon}), we find
\beq
 \dot{Q}_{\rm tot}(t) = -P_{\rm ac}(t) - P_{\rm el}(t).
 \edq
 This equation is the counterpart of Eq.~(\ref{delqtg1}), which has been derived within the thermodynamical approach for small changes in the equilibrium system. In the present case, it states that
 at every time, the power developed by the external sources, including the ac forces as well as the dc batteries that impose the chemical potential bias, is dissipated in the form of heat.  

On the other hand, it is interesting to notice that for any Hamiltonian $\calh_S(t)$ entering Eq.~(\ref{eq_H})
we can write the variation in time of the energy stored in the central part as
\beq\label{es}
 \dot{E}_S(t) \equiv \dot{\langle \calh_S \rangle} = J^E_S(t) - P_{\rm ac}(t),
\edq
which does not have a net contribution since $\overline{\dot{E}_S(t)}=0$.
Then, Eq. (\ref{prin1}) can also be expressed as
\ba\label{1stlaw}
& & \sum_{\alpha}  \left[J^E_{\alpha}(t)- \mu_{\alpha} \frac{I^C_{\alpha}(t)}{e} +J^E_{c\alpha}(t) \right]+ \dot{E}_S(t)-\mu \frac{I^C_S(t)}{e} \nonumber \\
& &\;\;\;\;\;\;\; + \; P_{\rm ac}(t)+P_{\rm el}(t)=0.
\ea

At this point it is important to stress that we have not made any assumption on the nature of the central system and on the characteristics of the driving. All the equations derived in this section rely on conservation laws only.

\subsection{Instantaneous heat fluxes through the different parts of the setup}
In Sec.~\ref{a} we have presented the definitions of the total heat and work fluxes consistent
with the thermodynamical approach of Sec.~\ref{thermo}.
As stressed before, these equations are exact and general. They do not rely on any particular method to evaluate the different fluxes or on the model 
describing the full setup. Equation~(\ref{1stlaw}) expresses the total heat produced at time $t$ in the full setup composed by the central structure, the reservoirs, and the contacts. The 
behavior of the time-average of the different fluxes in Eq.~(\ref{av1}) implies that 
\beq \label{qtotav}
\overline{\dot{Q}_{\rm tot}} =  \sum_{\alpha} \overline{\dot{Q}_{\alpha}} = - \overline{P_{\rm ac}}- \overline{P_{\rm el}},
\edq
with
\beq \label{qalfav}
 \overline{\dot{Q}_{\alpha}}  =  \overline{J^E_{\alpha}}- \mu_{\alpha} \frac{\overline{I^C_{\alpha}}}{e},
\edq
which is the usual definition of the dc-heat flux in the reservoir $\alpha$ ~\cite{defq}.  Equation~(\ref{qtotav}) reflects the fact  that the net heat production takes place at the reservoirs.

 In this section, we would like to discuss the role of the other terms entering Eq.~(\ref{1stlaw}), which do not contribute to the time-average but to the instantaneous
 total heat production. A possible interpretation of these terms is to identify them as components of the instantaneous 
  heat fluxes flowing through the different pieces of the device. 
Because of  the coupling between the central system and the reservoirs this interpretation is quite nontrivial, see e.g. Refs.~\cite{Dare:2016jk,Bruch:2015ux,Ankerhold:2014ky,Esposito:2015bg}.

Here we follow the approach introduced in Ref.~\cite{us}, where we considered the simple problem of a single driven level coupled to one reservoir and we argued that the appropriate definition of the  {\em time-dependent} heat current flowing into the reservoir $\alpha$ is
\beq\label{defheatres}
 \dot{Q}_{\alpha}(t)= J^E_{\alpha}(t) + \frac{J^E_{c\alpha}(t)}{2} - \mu_{\alpha} \frac{I^C_{\alpha}(t)}{e}.
\edq
Notice that, in addition to the terms contributing to the time-average given by Eq.~(\ref{qalfav}), we are adding half of the instantaneous rate of change of the energy stored at the contact [cf. the second term in the right-hand side of Eq.~\eqref{defheatres}].
 The arguments supporting this definition were presented in Ref.~\cite{us} and are the following: (i) it is consistent with the first law of thermodynamics, (ii) it matches the definition obtained in continuum models solved by scattering matrix formalism, and (iii) for the problem of an adiabatically driven level coupled to a single reservoir at zero temperature it leads to an instantaneous Joule-heating law, implying consistency with the second law of thermodynamics. The latter argument is worth of being highlighted. In fact, for a single driven system within the adiabatic regime in contact to a reservoir at $T=0$ we can just expect the
 heat flux to enter the reservoir at every time. The exact calculation presented in Ref.~\cite{us} shows that this is indeed the case when the definition given by Eq.~(\ref{defheatres}) is considered, whereas if the second term is not included in the definition of the instantaneous heat flux, we get the nonphysical result of a heat flux exiting a reservoir at zero temperature for some instants. Without the consideration of this term, no
 agreement can be obtained between the expressions of the scattering matrix formalism for continuum models and the ones derived with Green's function formalisms with discrete tunneling contact regions.
Finally, Ref.~\cite{guillem} shows that Eq.~\eqref{defheatres} leads to frequency-dependent heat current expressions that exhibit a proper parity property when the ac frequency is reversed.

In the case of a multiple-terminal setup, this definition of instantaneous heat flux through the reservoir $\alpha$ [Eq.~\eqref{defheatres}] is also in agreement with the scattering matrix one, as we show in detail in Appendix \ref{scat-mat}. Furthermore,  Eq.~(\ref{1stlaw}) suggests the following definition for the heat flux in the central piece of the system
\beq\label{defheats}
 \dot{Q}_{S}(t)= \dot{E}_S(t) - \mu \frac{I^C_{S}(t)}{e}+  \sum_{\alpha} \frac{J^E_{c\alpha}(t)}{2} .
\edq
We stress that $\mu$ is the chemical potential of the grounded reservoir. 
In this way,
\beq \label{qtot3}
\dot{Q}_{\rm tot}(t)= \sum_{\alpha} \dot{Q}_{\alpha}(t) +  \dot{Q}_{S}(t).
\edq
In Sec.~\ref{example} we analyze in more detail this splitting of the total rate of heat production for a concrete example. We will see that the interpretation of $\dot{Q}_{\alpha}(t)$ as the heat flux 
flowing into the reservoir and $\dot{Q}_{S}(t)$ as the one through the central system is, in fact, meaningful within the adiabatic regime for the driving and within linear response for the bias voltage.

\subsection{Instantaneous entropy production} \label{st}
As discussed in Sec.~\ref{c}, the power developed by the dissipative 
forces
is related to the heat and entropy production, while the power developed by the conservative forces leads to reversible heat with strictly zero average.
We define the conservative component of the force as
\beq\label{fcons}
{\bf F}^{\rm cons} (t)= -\mbox{Tr} \left[ \rho^{f} \frac{\partial {\cal H} }{ \partial {\bf V} } \right],
\edq
where $\rho^{f}$ is the frozen density operator, i.e., the equilibrium density operator considering the Hamiltonian ${\cal H}$ frozen at time $t$.
Hence, the instantaneous rate of entropy production reads
\ba \label{2law}
\dot{S}^{\rm rev}(t) &= & \frac{1}{T} \dot{Q}^{\rm rev}_{\rm tot}(t) = -\frac{1}{T} P^{\rm cons}_{\rm tot}(t),\nonumber \\
\dot{S}^{\rm diss}(t) &=& \frac{1}{T} \dot{Q}^{\rm diss}_{\rm tot}(t) = -\frac{1}{T} P^{\rm diss}_{\rm tot}(t),
\ea
with $P_{\rm tot}(t)=P_{\rm el}(t)+ P_{\rm ac}(t)= P^{\rm diss}_{\rm tot}(t)+  P^{\rm cons}_{\rm tot}(t)$.
Here we stress that the power developed by the batteries $P_{\rm el}(t)$ is only dissipative, while the power developed by the ac forces has dissipative and conservative components.
From the definition of the heat flux through the central system, Eq.~(\ref{defheats}), and the definition of the energy stored in this piece of the setup, Eq.~(\ref{es}), we can write the dissipative component of
this flux simply by subtracting the conservative component of the power. The result is
\beq \label{q}
\dot{Q}^{\rm diss}_S(t)=\dot{Q}_S(t) + P^{\rm cons}_{\rm tot}(t).
\edq

On the other hand, it is natural to conjecture that the heat production at the reservoirs is purely dissipative. 
Then, we express the irreversible entropy production as
\beq \label{2law-1}
\dot{S}^{\rm diss}(t)= \frac{1}{T}\left[\sum_{\alpha}  \dot{Q}_{\alpha}(t) + \dot{Q}^{\rm diss}_S(t) \right] = -\frac{1}{T} P^{\rm diss}_{\rm tot}(t).
\edq
As stressed in Sec.~\ref{c} the reversible component of the heat flux, related to the conservative forces contribute only dynamically. In fact, when averaging
over one cycle, the net contribution vanishes
\beq
\overline{P^{\rm cons}_{\rm tot}}= \overline{\dot{Q}^{\rm rev}_{\rm tot}}= \overline{\dot{S}^{\rm rev}}=0.
\edq
Instead, the dissipative entropy production $\dot{S}^{\rm diss}(t)$ has a non-vanishing average. This does not mean that all the terms of Eq.~(\ref{2law-1}) have a nonvanishing 
average. In fact, from the conservation laws Eqs.~(\ref{av1}) and~(\ref{av2}) we can see that 
\beq
\overline{\dot{Q}^{\rm diss}_S}=\overline{\dot{Q}_S}=0
\edq
and also the terms $J^E_{c\alpha}(t)$ entering $\dot{Q}_{\alpha}(t)$ have a zero average, as discussed in Ref.~\cite{us}. 
 In the next section, we will further analyze the role of these terms. We anticipate that they are crucial to guarantee the second law instantaneously, in the sense that
at each time

 \beq \label{sdis}
 \dot{S}^{\rm diss}(t) \geq 0.
 \edq

\section{Time-dependent adiabatic approach} \label{tdadia}
In this section we focus on slow driving. Our analysis will be based on the approach presented in Ref.~\onlinecite{thermolud}, which consists of a linear response picture 
akin to Kubo theory in $\delta \mu_{\alpha}$ combined to an adiabatic expansion in $\dot{\bf {V}}$. For the sake of clarity, we consider a two-terminal setup with left and right reservoirs,
$\alpha=L,R$, and $\mu_L=\mu$ and $\mu_R=\mu-\delta\mu$. In this approach, the forces and the currents, as well as the mean value of any observable, is regarded as an expansion
in powers of $\delta \mu, \dot{V}$.  In what follows, we focus on the forces and the charge current entering the right reservoir, and keep up to linear order in these parameters:
\ba
F_j(t)& = &  F_j^{\rm cons}+ \sum_l \Lambda^{FV}_{jl} \dot{V}_l+\Lambda^{F\mu}_{j} \delta \mu, \nonumber \\
I_R^C(t) & = & \sum_l \Lambda^{C V}_{l} \dot{V}_l+\Lambda^{C \mu} \delta \mu,
\ea
where $F_j^{\rm cons}$ was defined in Eq.~(\ref{fcons}) and the linear response coefficients are related to susceptibilities evaluated with the frozen density operator 
$\rho^{f}$. Their dependence on time is calculated from the frozen Hamiltonian evaluated at $t$ \cite{thermolud}. Hence, the power developed by the ac forces and by the dc batteries read, respectively,
\ba \label{p}
P_{\rm ac}(t) & = & P_{\rm ac}^{\rm cons}(t)+  \sum_{j l} \Lambda^{FV}_{jl} \dot{V}_l(t) \dot{V}_j(t)  + \sum_j \Lambda^{F\mu}_{j} \delta \mu \dot{V}_j(t), \nonumber \\
P_{\rm el}(t) & = & - \sum_l \Lambda^{C V}_{l} \dot{V}_l \delta \mu - \Lambda^{C \mu} \delta \mu^2,
\ea
with
\beq
P_{\rm ac}^{\rm cons}(t)= P_{\rm tot}^{\rm cons}(t)= \sum_j   F_j^{\rm cons} \dot{V}_j(t).
\edq
In Eq.~(\ref{p}) the negative sign of $P_{\rm el}$ follows the definition given by Eq.~(\ref{pva}). 
As shown in Ref.~\cite{thermolud} for systems with time-reversal symmetry, the coefficients $\Lambda$ obey microreversibility and satisfy Onsager relations
\beq
 \Lambda^{FV}_{jl}  =  \Lambda^{FV}_{lj}, \;\;\;\;\; \Lambda^{C V}_{l}=  \Lambda^{F\mu}_{l}.
\edq
Therefore the instantaneous dissipated power defining the rate of entropy production is
\beq \label{pdis}
P^{\rm diss}_{\rm tot}(t)= \sum_{j l} \Lambda^{FV}_{jl} \dot{V}_l(t) \dot{V}_j(t) - \Lambda^{C \mu} \delta \mu^2.
\edq
This term must be positive in order to satisfy the instantaneous second law, Eq.~(\ref{sdis}).

\section{Example. A single driven level coupled to two reservoirs} \label{example}
In order to analyze the theoretical concepts introduced above, we consider a simple central system of the form  (see sketch of Fig.~\ref{fig1})
\beq
\calh_S=\veps_d(t)d^\dagger d,
\edq
 which consists of a driven single resonant energy level 
$\veps_d(t)=\veps_0+V(t)$ 
(e.g., a quantum dot) coupled to two fermionic baths (left and right), with $\mu_L=\mu$ and $\mu_R=\mu-\delta\mu$, respectively. Both of them are kept at the same temperature, $T$. 
In Ref.~\cite{us} we considered the single reservoir case with $T=0$, and now we extend the configuration to multiple reservoirs and to finite temperature.

\subsection{Green's function approach} \label{green}
The different currents and energy fluxes, as well as the power developed by the ac forces, can be computed in terms of the retarded Green function $G^R(t,t')=-i\theta(t-t')\langle\{d(t),d^\dagger (t')\}\rangle$ and the lesser Green function $G^<(t,t')=i\langle d^\dagger (t')d(t)\rangle$ of the central structure, which can be obtained by solving a Dyson equation~\cite{lili1,lili2,lilimoskalets}. 

To compute the time-dependent heat current entering the reservoir $\alpha$ given by Eq.~(\ref{defheatres}), we need to have an expression for the charge current $I^C_\alpha$, the energy currents $J^E_\alpha$ and $J^E_{c\alpha}$. Generalizing Ref.~\onlinecite{us} to the case of many reservoirs, we start by performing the Fourier transform of the Green function
\ba
&&G^{R,<}(t,t^{\prime})=\int_{-\infty}^{\infty}\frac{d\veps}{2\pi}e^{-i\frac{\veps}{\hbar}(t-t')}G^{R,<}(t,\veps),  \\
&&G^<(t,t^{\prime}) = \sum_{\alpha} \int_{-\infty}^{\infty}\frac{d\veps}{2\pi}e^{-i\frac{\veps}{\hbar}(t-t')}G^{R}(t,\veps)\Sigma^<_{\alpha}(\veps)[G^{R}(t^{\prime},\veps)]^*,
\ea
where $\Sigma^<_{\alpha}(\veps)= i f_{\alpha}(\veps) \Gamma_{\alpha}$.  We have  introduced the hybridization with the reservoir $\alpha$, $\Gamma_\alpha=\sum_{k_\alpha}2\pi\vert w_{k_\alpha}\vert^{2}\delta (\veps-\veps_{k_\alpha})$ and  $f_\alpha(\veps)=[e^{(\veps-\mu_\alpha)/(k_BT)}+1]^{-1}$ is the Fermi-Dirac distribution of the reservoir labeled with $\alpha$. In the case of  reservoirs with a wide band, 
in which $\Gamma_\alpha$ is a constant function, 
the charge current flowing into lead $\alpha$ reads

\beq\label{charge1}
I^{C}_{\alpha}(t)=-\frac{e}{h}\int d\veps \,\Gamma_\alpha\,2\mbox{Re}\{iG^{R}(t,\veps)f_\alpha(\veps)+G^{<}(t,\veps)\Theta(\veps)\},
\edq
where $\Theta(\veps)=\int \frac{d\veps'}{2\pi}\frac{1}{\veps-\veps'-i0^{+}}$, and the energy current entering reservoir $\alpha$ is
\beq\label{energy1}
J^{E}_{\alpha}(t)=-\int \frac{d\veps}{h} \,\Gamma_\alpha\,2\mbox{Re}\{i G^{R}(t,\veps)f_\alpha(\veps)\veps +G^{<}(t,\veps)\theta(\veps)\},
\edq
with $\theta(\veps)=\int \frac{d\veps'}{2\pi}\frac{\veps'}{\veps-\veps'-i0^{+}}$.

On the other hand, the variation of the energy stored in the contact region between the central system and the reservoir $\alpha$ can be written as
\beq
J^{E}_{c\alpha}(t)=\int\frac{d\veps}{2\pi}\Gamma_\alpha \,f_\alpha(\veps)2\,\mbox{Re}\{\partial_t G^{R}(t,\veps)\},
\edq
and the power performed by the ac potentials is
\beq
P_{\rm ac}(t)=\dot{V}(t)\int\frac{d\veps}{2\pi}\,\mbox{Im}\{G^{<}(t,\veps)\}.
\edq

Now, taking into account that the ac external potential is periodic in time, it is convenient to introduce the Floquet-Fourier representation for the Green function \cite{lili1,lili2} 
\beq
G^{R}(t,\veps)=\sum_{n=-\infty}^{\infty}e^{-in\omega t}\calg(n,\veps),
\edq
where $\omega=2\pi/\tau$ is the oscillation frequency of the ac parameter $V(t)$. Using this representation, the charge current entering reservoir $\alpha$ reads 
\ba
\label{charge}
&I^C_{\alpha}(t) & =\frac{e}{h}\sum_l \int {d\veps} e^{-i l \omega t}   {\Gamma}_\alpha 
\{   
i {\cal G}^{*}(-l,\veps) 
 \left[f_\alpha(\veps) -f_\alpha(\veps_l)\right] - \nonumber \\
& & \sum_{n,\beta} 
\left[f_\alpha(\veps)- f_\beta(\veps_n)\right]{\Gamma}_\beta
  {\cal G}(l+n,\veps_n) 
{\cal G}^{*}(n,\veps_n)  \},
\ea
with $\beta=L,R$ and $\veps_n=\veps -n\hbar\omega$. On the other hand, the energy current flowing into $\alpha$ is
\ba
\label{energy}
& J^E_\alpha (t) &=\sum_l\int \frac{d\veps}{h}  e^{-i l \omega t}  {\Gamma}_\alpha  
\{   
i {\cal G}^{*}(-l,\veps) 
 \left[\veps f_\alpha(\veps) -\veps_l f_\alpha(\veps_l)\right] \\
& & -\sum_{n,\beta} 
\left[\veps f_\alpha(\veps)-\veps_{-\frac{l}{2}} f_\beta(\veps_n)\right]{\Gamma}_\beta 
  {\cal G}(l+n,\veps_n)
{\cal G}^{*}(n,\veps_n) \},\nonumber 
\ea
and the variation of the energy corresponding to the contact region can be written as
\beq
 J^E_{c\alpha} (t)=\int\frac{d\veps}{h} f_\alpha(\veps)\Gamma_\alpha\sum_l l\hbar\omega\,2\mbox{Im}\{{\cal G}(l,\veps)e^{-il\omega t}\}.
\edq

Then, the time-dependent heat flux $\dot{Q}_\alpha(t)$ of Eq.~(\ref{defheatres}) reads
\ba\label{heat}
&& \dot{Q}_{\alpha}(t) =\sum_l \int \frac{d\veps}{h}  e^{-i l \omega t} {\Gamma}_\alpha 
\{   
i {\cal G}^{*}(-l,\veps)(\veps_{\frac{l}{2}}-\mu_\alpha)\nonumber\\
 && \times\left[f_\alpha(\veps) -f_\alpha(\veps_l)\right] \nonumber \\
& &- \sum_{n,\beta} 
(\veps_{-\frac{l}{2}}-\mu_\alpha)\left[f_\alpha(\veps)- f_\beta(\veps_n)\right]{\Gamma}_\beta
  {\cal G}(l+n,\veps_n) 
{\cal G}^{*}(n,\veps_n)  \},\nonumber\\
&& 
\ea

In Ref.~\onlinecite{us} we have demonstrated the equivalence  between this expression and the one derived within scattering matrix formalism for the case of a singe reservoir.
In Appendix \ref{scat-mat} we show that the definition given by Eq.~(\ref{defheatres}), expressed in terms of Green's functions in Eq. (\ref{heat}) for the more general case of multiple reservoirs, is also in agreement with the expression for the heat current calculated derived within scattering matrix theory.

Similarly, the power performed by the ac potentials is
\ba\label{powergf}
P_{\rm ac}(t)& = & \sum_\alpha\sum_{l,m,n}\int\frac{d\veps}{h}n\hbar\omega f_\alpha(\veps){\Gamma}_\alpha\\\nonumber
& & \times\mbox{Im}\{{V}(n){\cal G}(m+l,\veps){\cal G}^{*}(l,\veps)e^{-i\omega t(m-n)}\},
\ea 
where $V(n)$ are the Fourier components of $V(t)=\sum_n V(n)e^{in\omega t }$.

\subsection{Heat flow in the adiabatic regime}
In the adiabatic regime, we rely on the expansion in powers of $\dot{\bf V}$ and $\delta \mu$ presented in Sec.~\ref{tdadia}. In order to evaluate the coefficients $\Lambda$ for this specific problem, 
we start from the expressions of the power and the currents given in Section \ref{green} and perform an expansion up to second and linear order, respectively in $\omega$ and
$\delta \mu$ (notice that $\dot{V} \propto \omega$ in the present problem). From these expansions (see Appendix \ref{cq} ), we can directly identify the coefficients
$\Lambda$. The explicit expressions are shown in Appendix \ref{coefficients}. These coefficients depend on  the frozen density of states (or spectral function) $\rho^{f}$, with the time $t$ treated as a parameter. 

\begin{figure}[b]
  \includegraphics[scale=0.32]{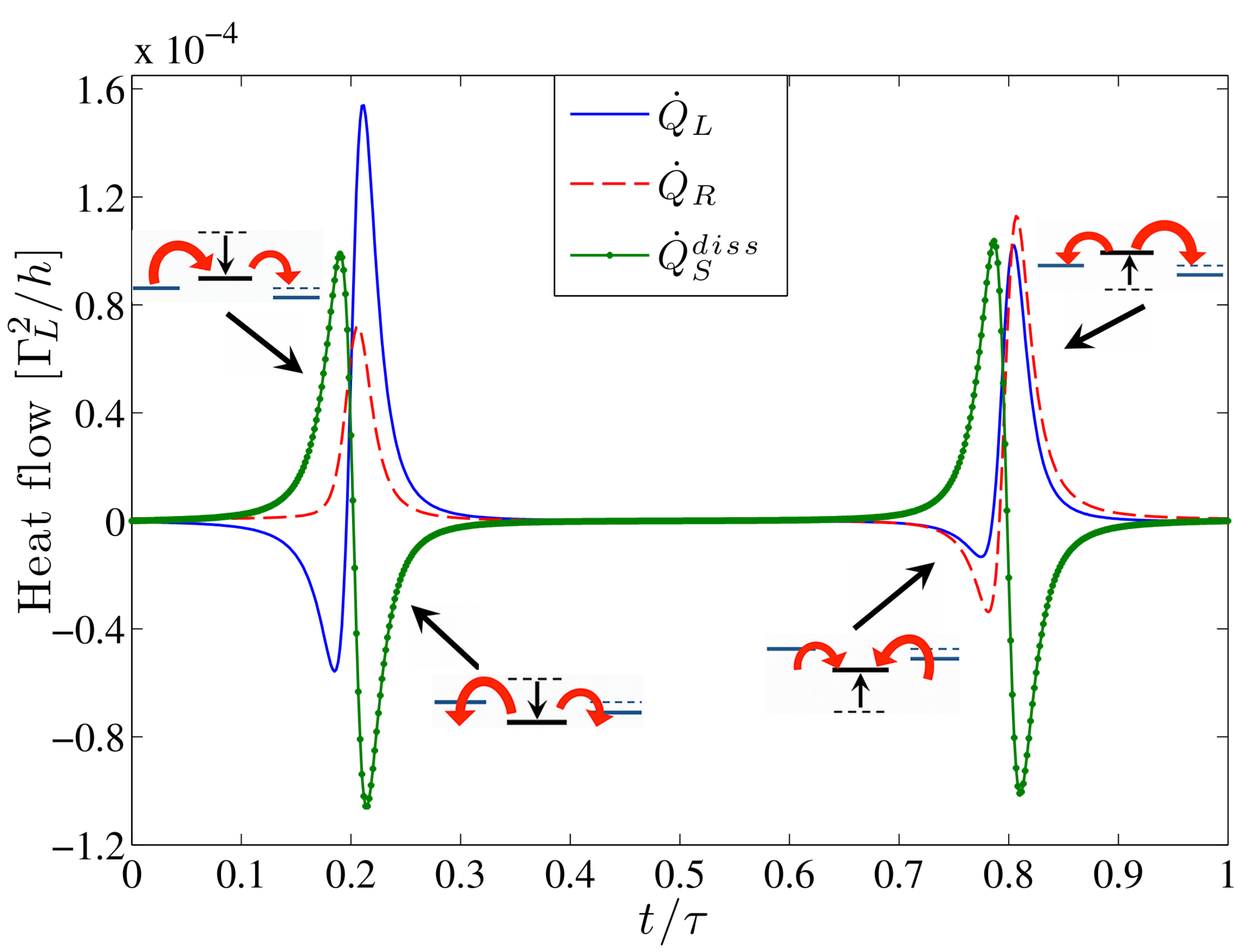}
  \caption{Heat fluxes at the left (solid lines) and right (dashed lines) reservoirs as well as the flux  $\dot{Q}_S^{\rm diss}(t)$ (circles) for a driven level connected to reservoirs at finite temperature $k_B T=0.05$ and with a small applied bias $\delta\mu=0.004$. The energy of the level evolves in time as $V(t)=7\cos(\omega t)$ with $\hbar\omega=10^{-3}$. Parameters: $\mu=2$, $\veps_0=0$, and the hybridization are $\Gamma_L=1$ and $\Gamma_R=0.6$. Energies are expressed in units of $\Gamma_L$.
 Sketches illustrating the physical processes as function of time are also provided. In each case, the horizontal central  line indicates the position of the level at a given time referred to the position of the chemical potentials of the reservoirs, while the red arrows indicate the direction of the heat flux associated to the reservoirs.}\label{f-heat}
\end{figure}

In particular, starting from  Eq.~(\ref{heat}) to compute $\dot{Q}_\alpha(t)$ up to second order in $\omega, \delta \mu$ we find
 ${\dot{Q}_\alpha}(t)=\dot{Q}_\alpha(t)^{(1)}+\dot{Q}_\alpha(t)^{(2)}$ with 
\ba\label{heatres}
\dot{Q}_\alpha(t)^{(1)} &=& \Lambda_\alpha^{\dot{V}} \,\,\dot{V}+\Lambda_\alpha^{\delta\mu} \,\,\delta\mu,\\
\dot{Q}_\alpha(t)^{(2)} &=& \Lambda_\alpha^{\dot{V}^2} \,{\dot{V}^2}+\Lambda_\alpha^{\ddot{V}} \,\,{\ddot{V}}+\Lambda_\alpha^{\dot{V}\delta\mu} \,\,{\dot{V}\delta\mu}+\Lambda_\alpha^{{\delta\mu}^2} \,{{\delta\mu}^2}.
\ea
On the other hand, if we take into account the relation for the entropy production of Eq.~(\ref{2law-1}), and the expressions within the low frequency approximation,  Eqs.~(\ref{heatres}), (\ref{pdis}) and (\ref{lambdas}), we can also compute $\dot{Q}^{\rm diss}_S(t)$ up to second order in $\omega, \delta \mu$
as ${\dot{Q}^{\rm diss}_S}(t)=\dot{Q}^{\rm diss}_S(t)^{(1)}+\dot{Q}^{\rm diss}_S(t)^{(2)}$, where the first and second order are 
\ba \label{qsdis}
\dot{Q}^{\rm diss}_S(t)^{(1)} &=&\Lambda_S^{\dot{V}}\,{\dot{V}},\nonumber\\
\dot{Q}^{\rm diss}_S(t)^{(2)} &=&\Lambda_S^{{\dot{V}}^2}\,{{\dot{V}}^2}+\Lambda_S^{\ddot{V}}\,{\ddot{V}}+\Lambda_S^{\dot{V}\delta\mu}\,{\dot{V}}\delta\mu.
\ea
The behavior of the heat flux at the two reservoirs, along with $\dot{Q}^{\rm diss}_S(t)$ within a period, is shown in Fig. \ref{f-heat} for reservoirs at finite temperature $T$ and a small applied bias voltage $\mu_L-\mu_R=eV$. 
For $t=0$ the energy of  the level is above the highest chemical potential $\mu_L$. As $t$ evolves, the energy of level approaches  $\mu_L$ from above and when $ \veps(t) -\mu_L \sim k_B T$ a heat flux leaves the left reservoir, traveling through the central level towards the right reservoir.  When the energy of the level becomes approximately aligned with the mean chemical potential of the reservoirs, the heat flow
goes from the central piece into the two reservoirs.  Later, the level lies well below the lowest chemical potential $\mu_R$ and the heat flux becomes vanishingly small. When the level oscillation completes half a period ($t=\tau/2$), the motion reverses and approaches $\mu_R$ from below. For $\mu_R - \veps(t) \sim k_B T$, a heat flux is established from the reservoirs to the central piece until the level aligns with the mean chemical potential. Then, the heat flows from the central system into the reservoirs. 

\begin{figure}[b]
  \includegraphics[scale=0.55]{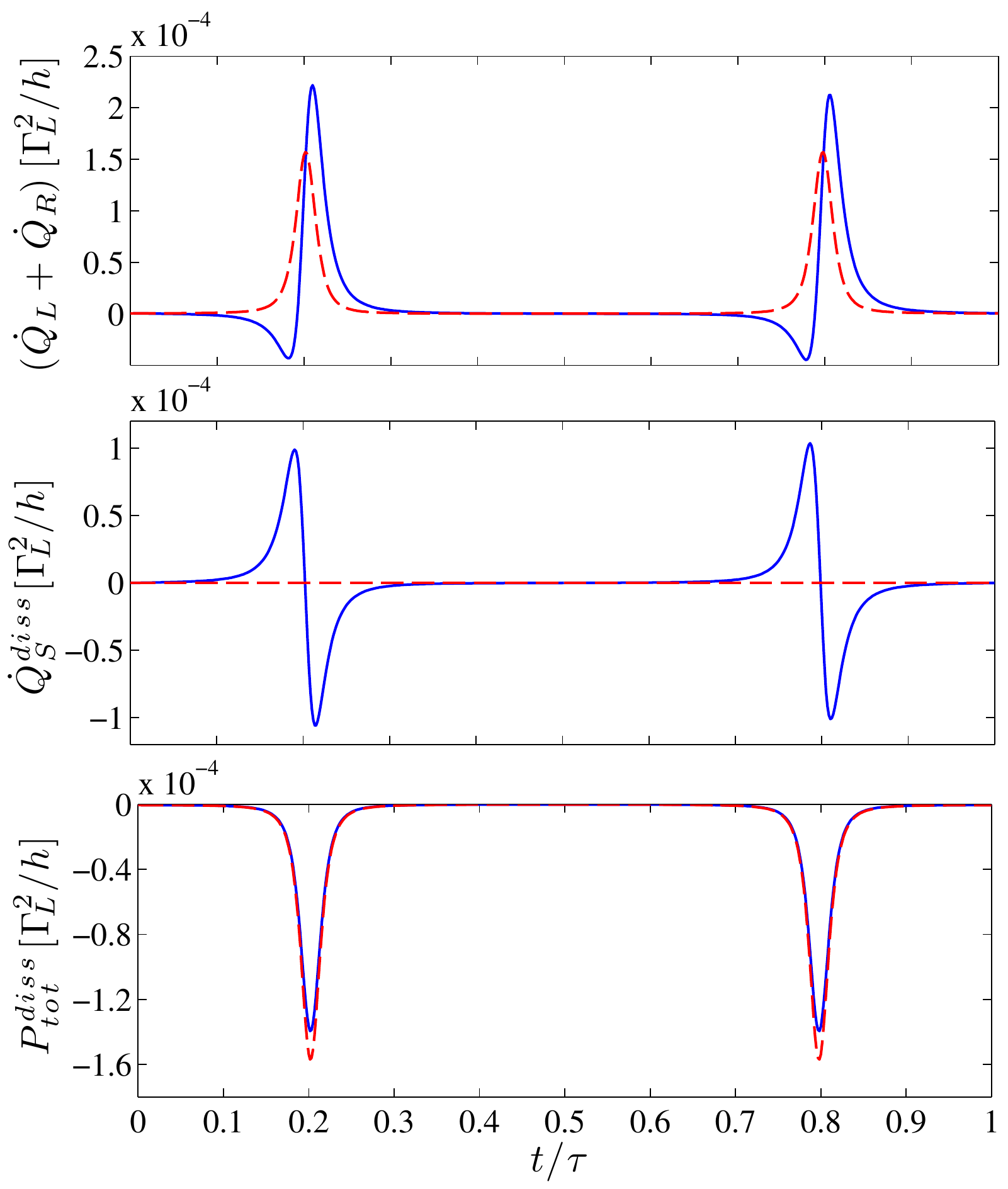}
  \caption{The different components of the total heat production $\dot{Q}_{\rm tot}^{\rm diss}(t)= -P^{\rm diss}_{\rm tot}(t)$ as a function of time for a single level coupled to two reservoirs within the adiabatic regime. Dashed lines corresponds to reservoirs at $T=0$, while solid lines are for $k_BT=0.05$. Other parameters are the same as in Fig. \ref{f-heat}. Energies are expressed in units of $\Gamma_L$. The upper panel shows that the heat flux at the reservoirs is positive and equal to $-P^{\rm diss}_{\rm tot}$ at $T=0$ and may attain negative values at finite temperature. The dissipative heat flux at the driven dot $\dot{Q}_S^{\rm diss}(t)$, in the second panel, vanishes when $T=0$. The bottom panel shows the total dissipative power $P^{\rm diss}_{\rm tot}$.  
}\label{fig3}
\end{figure}

\begin{figure}[!ht]
  \includegraphics[scale=0.33]{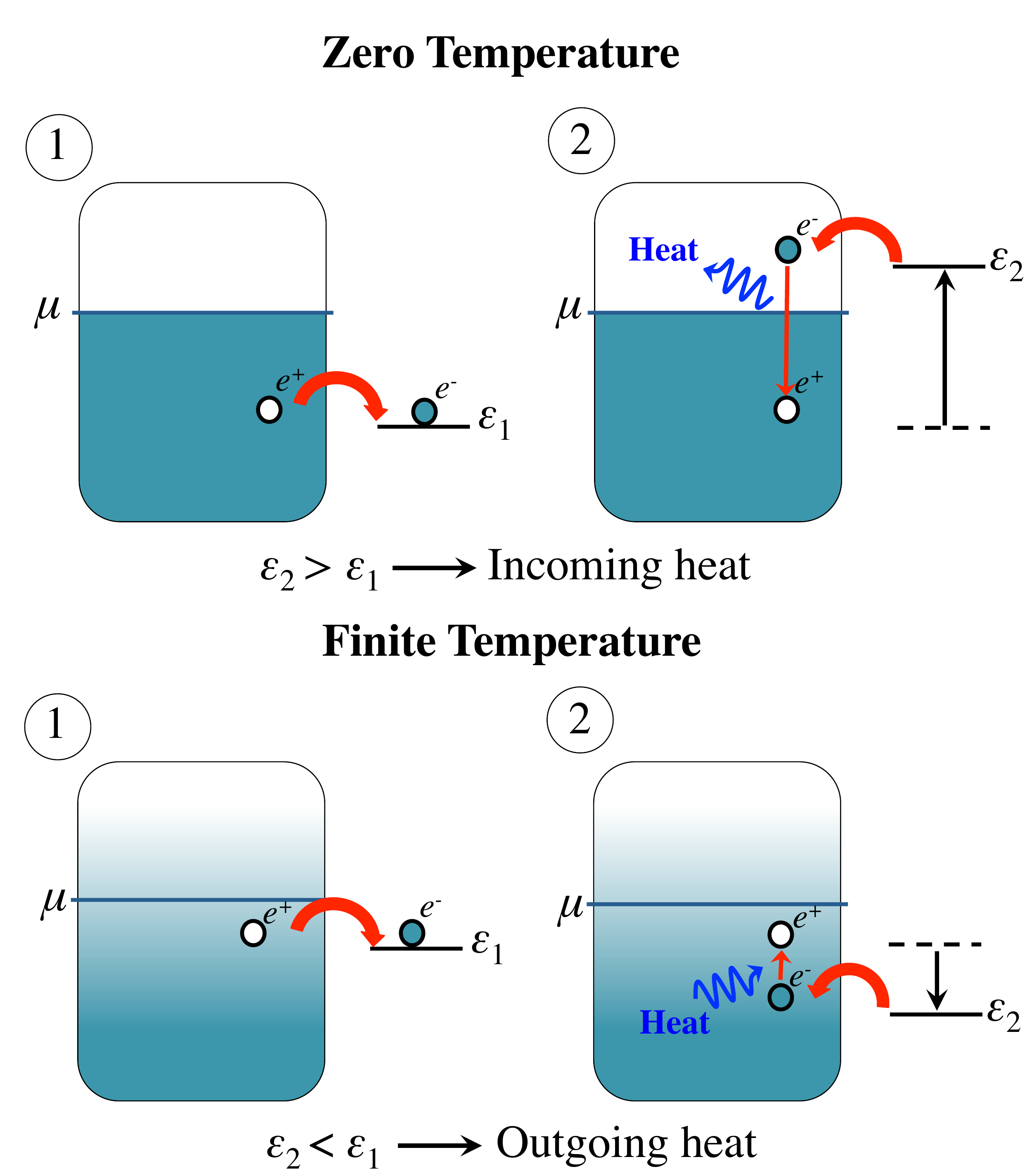}
  \caption{Sketches illustrating the heat exchange between the reservoirs and the central piece. The upper panel corresponds to $T=0$, in which case the heat generated by the driving can only be injected into the reservoir. The lower panel corresponds to finite temperature and indicates that depending on the position of the level relative to the chemical potential of the reservoir, heat flow can be inwards or outwards.}\label{fig2}
\end{figure}

It is interesting to analyze the total entropy production of the above processes as a function of time. Let us start by 
noticing that $\rho^f_\alpha\ge0$ and $\partial_\veps f\le0$ . Then, from Eqs.~(\ref{pdis}) and (\ref{lambdas}) for the dissipated power in the adiabatic regime, it follows that $P^{\rm diss}_{\rm tot}(t)\le0$ and therefore 
\beq\label{enpos}
\dot{S}^{\rm diss}(t)\ge 0.
\edq
As discussed in Sec.~\ref{st}, the instantaneous rate of entropy production contains terms associated to the heat production in the reservoirs as well as terms
associated to the heat production at the central piece, as explicitly defined in Eq.~\eqref{2law-1}. 
While in Fig.~\ref{f-heat} each of these contributions is  separately analyzed, in Fig.~\ref{fig3} we show the combined effect. 
Interestingly, $\dot{Q}_{\rm tot}^{\rm diss} (t) =0$ for $T=0$, which can be exactly verified by noticing that the coefficients $\Lambda_S$ entering (\ref{qsdis}) contain integrands with 
 $ (\veps -\mu)\partial_\veps f  = - (\veps -\mu)\delta (\veps - \mu)$ at $T=0$. The physical explanation to this property is the fact that for $T=0$ all the dissipation takes
place at the reservoirs. In fact, for reservoirs at zero temperature, heat can only be injected from the central system into the reservoirs, which means that
$\dot{Q}_{\rm tot}^{\rm diss} (t) = \sum_{\alpha} \dot{Q}_{\alpha} (t) \geq 0$. However, at finite temperature, the reservoirs can be temporarily cooled down  as shown in Fig. \ref{fig2} and we could have
$\sum_{\alpha} \dot{Q}_{\alpha} (t) \leq 0$. In that case, the only possibility to have $\dot{Q}_{\rm tot}^{\rm diss} (t) \geq 0$ is to have a positive non-vanishing 
${\dot{Q}^{\rm diss}_S}(t) \neq 0$. This is illustrated in Fig. \ref{fig3}, where the behavior of the total dissipated power is also shown.

\subsection{Instantaneous Joule law in the adiabatic regime}
In Ref.~\cite{us} we showed that the heat production by a single driven dot connected to a single reservoir at $T=0$ obeys an instantaneous Joule Law in the adiabatic regime. 

The corresponding resistance is universal and equal to the charge relaxation resistance  $R_q= h/2e^2$ introduced in Ref.~\onlinecite{Buttiker:1993bw} and observed in Ref.~\onlinecite{Gabelli:2006eg}.
We are interested now in analyzing a possible relation in the case of a dot connected to two reservoirs that may have a finite temperature and a bias voltage. We rely again on linear response. 
The term $\Lambda^{F,V} \dot{V}^2$ describes the heat dissipated due to the variation in time of the ac potential, and $\Lambda^{C,\mu}\delta\mu^2$ captures the effect of the applied static bias $\delta\mu$.
The first term, which is proportional to $\dot{V}^{2}$, can be expressed in a different way by evaluating the charge current entering the system $I^{C}_S$ up to first order in the velocity $\dot{V}$ as
\beq\label{isj}
{I^{C}_S}^{(1)}(t)=-\sum_\alpha{{I^{C}_\alpha}^{(1)}(t)}=e\int \frac{d\veps}{2\pi}\partial_\veps f\rho^{f}\dot{V}.
\edq
For this, we used the relation given by Eq.~(\ref{numbal2}) and the expression for the currents entering the reservoirs, Eq.~(\ref{charge}), and follow the steps presented in Appendix~\ref{cq} for the slow driving case. Now, combining Eq.~(\ref{isj}) with the first term of Eq.~(\ref{pdis}) we get
\beq\label{p2}
-\Lambda^{F,V} \dot{V}^2=R_{\rm ac}(t)[{{I^{C}_S}^{(1)}(t)}]^{2},
\edq
where we have defined the resistance
\beq
R_{\rm ac}(t)=-\frac{h}{2e^{2}}\frac{\int{d\veps}\partial_\veps f (\rho^{f}(t,\veps))^{2}}{\left(\int{d\veps}\partial_\veps f \rho^{f}(t,\veps)\right)^{2}},
\edq
which is a manifestly positive quantity at all times. 
Therefore, we find that the heat dissipated due to pumping is given by a Joule law with an instantaneous resistance $R_{\rm ac}(t)$. 
This quantity becomes nonuniversal at finite temperatures, which agrees with the finite-temperature result of  Ref.~\cite{Buttiker:1993bw}.

\begin{figure}[t]
  \includegraphics[scale=0.52]{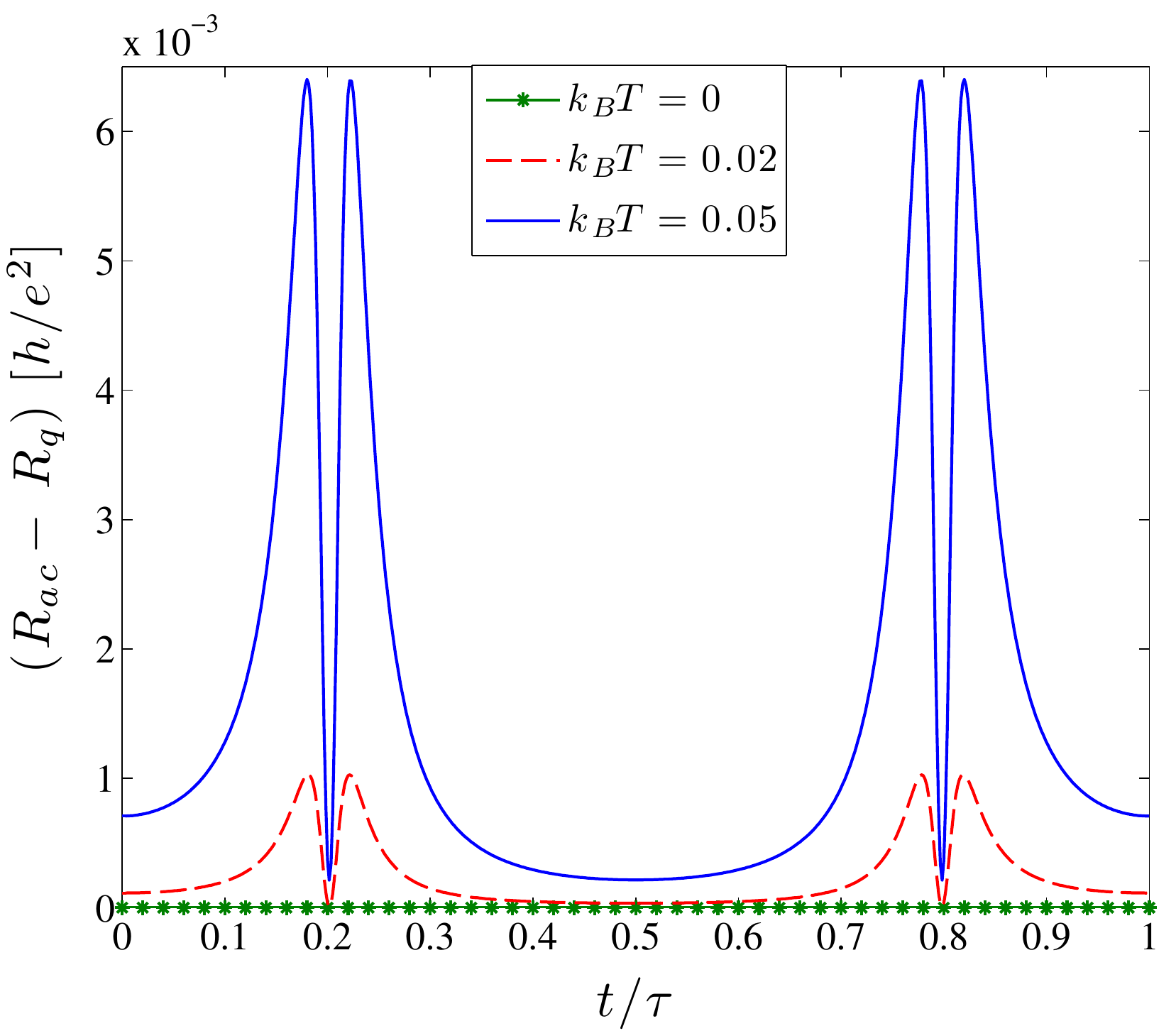}
  \caption{Difference between the ac resistance $R_{ac}(t)$ and the relaxation resistance quantum $R_q$ as a function of time within the adiabatic regime, and for different temperatures $T$. Other parameters are the same as in Fig.~\ref{f-heat}. Our results show that the instantaneous ac resistance becomes universal only at $T=0$, attaining the value $R_q$. For finite temperatures, we have $R_{ac}(t)>R_q$, hence the difference between the two is always a positive quantity.}\label{fig4}
\end{figure}

If the temperatures of the reservoirs are small compared to their Fermi energy, it is possible to apply the Sommerfeld expansion up to order $T^{2}$. Accordingly, we investigate the behavior of $R_{\rm ac}(t)$ at finite temperature 
\beq\label{rr}
R_{\rm ac}(t)\sim \frac{h}{2e^{2}}  \left. \left(1+ \frac{\pi^2 T^2 }{3}  \frac{( \partial_{\veps} \rho^{f})^{2} }{ (\rho^{f})^{2} } \right) \right\vert_{\veps=\mu}.
\edq
Remarkably, the resistance becomes universal at $T=0$, recovering the quantum of charge relaxation resistance $R_{\rm ac}^{T=0}=R_q=h/2e^{2}$. For low but finite 
temperatures, the resistance increases as shown in Eq.~(\ref{rr}) and becomes  $R_{\rm ac}^{T}>R_q$.  This is illustrated in Fig.~\ref{fig4}. The departures from the ideal quantum limit of the resistance are 
$\propto T^2$ and are sizable for those times when the energy of the level differs from the mean chemical potential of the reservoirs in an amount $\sim k_B T$.

On the other hand, for the quadratic term in the bias drop $\Lambda^{C\mu}\delta\mu^2$ of Eq. (\ref{pdis}), we also have an instantaneous Joule law of the form 
\beq
\Lambda^{C \mu }\delta\mu^2=\frac{[{I_R^{C}}(t)]^{2}}{G(t)},
\edq
with an electrical conductance $G(t)=\Lambda^{C ,\mu }$.

\section{Conclusions}
We have analyzed the dynamics of the energy transport and entropy production in an electron system coupled to multiple reservoirs and slightly driven out of equilibrium by means of ac local and dc bias voltages. 
We have formulated an exact  quantum dynamical approach, which allows to identify time-resolved quantities, such as  the total heat dissipated by the system, the work done on a system, and the entropy production, in a way which is fully consistent with the first and the second laws of thermodynamics. 

In addition, we identified conservative and dissipative contributions to the total heat flux and the corresponding contributions to the entropy production. 
The  time-resolved heat fluxes flowing through the different pieces of the device were investigated in detail. 
We have shown that the definition of the time-resolved heat current flowing into the reservoirs recently introduced in Ref.~\cite{us} for a single-terminal system is also suitable for multiterminal devices. 
This definition takes into account the energy temporarily stored in the contact region connecting the driven central system and the reservoirs. 
Using this definition, we showed that in the limit of a slow driving the first and the second laws of thermodynamics can be formulated consistently at each instant of time. 

We illustrated our approach by considering a simple example---a slowly driven resonant level coupled to two electron reservoirs at a finite temperature and with an applied bias voltage. 
We showed that at finite temperatures, when one of the reservoirs can be temporarily cooled, the total heat production at each time is positive, hence the entropy production is positive, only if the energy stored in the contact and central regions are taken into account.  Since all the equations of Sec.~III, IV and V have been derived under very general assumptions regarding the nature of the specific model, we
expect that the qualitative features in the behavior of the entropy production and instantaneous heat flow presented in Sec.~VI will remain valid even for  different types of periodic driving and also
the case of a quantum dot with many-body interactions.
The latter type of interactions may, however, affect other more quantitative features such as the behavior of the instantaneous Joule law analyzed in Section VI.C, since an electron-electron interaction is shown to affect the charge relaxation resistance $R_q$ at finite temperatures \cite{Nigg:2006kl}, for large cavities \cite{Mora:2010hw} and at finite magnetic fields \cite{Filippone:2011fy}.

Our results thus represent a significant advance toward a full understanding of dissipation and dynamics in quantum electron systems and might have important implications for nanoelectronics and quantum thermodynamics.

\section{Acknowledgements} 
This work was supported by MINECO under Grant No.~FIS2014-52564, UBACyT, CONICET and MINCyT, Argentina.

\appendix

\section{Relation to scattering matrix} \label{scat-mat}

Within the scattering matrix approach~\cite{Buttiker:1992ge,mbook}, the heat flux in the lead $\alpha$ reads
(see, e.g., Refs.~\cite{Battista:2013ew,Lim:2013de,Battista:2014di})
\begin{eqnarray}\label{heatscatt}
& &\dot{Q}^{S-M}_{\alpha}(t)    =  
\sum\limits_{l=-\infty} ^{\infty}e^{- il\omega t}  
\int \dfrac{d\veps}{h} \left(\veps_{-\frac{l}{2}} - \mu _{ \alpha} \right) 
\sum\limits_{n=-\infty}^{\infty} 
\sum\limits _{\beta = L,R}\nonumber\\
&&\times 
\left\{ f _{ \beta}\left(\veps_{n} \right) - f _{ \alpha}(\veps) \right\}  S _{ \alpha \beta}^{*}\left(\veps,\, \veps_{n} \right)\, S _{ \alpha \beta}\left(\veps_{-l},\, \veps_{n} \right) 
, 
\end{eqnarray}
where  $S(\veps_m,\veps_n)$ is the Floquet scattering matrix which is related to the Green function via the generalized Fisher-Lee relation \cite{lilimoskalets,Fisher-Lee}
\beq
S_{\alpha,\beta}(\veps_{-m},\veps_{-n})=\delta_{\alpha,\beta}\delta_{m,n}-i\sqrt{\Gamma_\alpha\Gamma_\beta}\calg(m-n,\veps_{-n}).
\edq 

From this relation we find that
\ba\label{ss}
 &&S _{ \alpha \beta}^{*}\left(\veps,\, \veps_{n} \right)\, S _{ \alpha \beta}\left(\veps_{-l},\, \veps_{n} \right)=i\,\delta_{\alpha\beta}\delta_{l,-n}\sqrt{\Gamma_\alpha\Gamma_\beta}\calg^*(n,\veps_{n})\nonumber\\
&&+\delta_{\alpha\beta}\delta_{n,0}
\times\left[\delta_{l,-n}-i\sqrt{\Gamma_\alpha\Gamma_\beta}\calg(l+n,\veps_{n})\right]\nonumber\\
&&+\Gamma_\alpha\Gamma_\beta\calg^*(n,\veps_{n})\calg(l+n,\veps_{n}),
\ea
and therefore Eq. (\ref{heatscatt}) reads 
\ba\label{heat2}
&& \dot{Q}_{\alpha}^{S-M}(t) =\sum_{l,n} \int \frac{d\veps}{h}  e^{-i l \omega t}  
 (\veps_{-\frac{l}{2}}-\mu_\alpha)\sum_\beta\left[f_\beta(\veps_n) -f_\alpha(\veps)\right] \nonumber \\
& &\times\calg^*(n,\veps_{n})\{i\,\delta_{\alpha\beta}\delta_{l,-n}\sqrt{\Gamma_\alpha\Gamma_\beta}+\Gamma_\alpha\Gamma_\beta\calg(l+n,\veps_{n})\}. 
\ea

Here, the term in Eq. (\ref{ss}) which is accompanied by $\delta_{\alpha\beta}\delta_{n,0}$ does not contribute due to the difference between the Fermi functions.
Then, after some algebra and by comparing with Eq. (\ref{heat}) we find
\beq
\dot{Q}^{S-M}_{\alpha}(t) =\dot{Q}_{\alpha}(t),
\edq
where $\dot{Q}_{\alpha}(t)$ is given in Eq. (\ref{defheatres}) and includes in the definition the contributions $J^E_{c\alpha}(t)$ due to the contacts.

\section{Slow driving and small bias voltage}\label{cq}

To calculate up to $\omega^{2}$ and $\delta\mu^{2}$ the currents (\ref{charge}), (\ref{heat}), and the power (\ref{powergf}), we need to perform an expansion of the Fermi function entering the integrands as
\beq
f_\alpha(\veps+n\hbar\omega)\sim f_\alpha(\veps)+{\partial_\veps f_\alpha}\hbar n\omega +\partial_\veps^{2}f_\alpha \frac{(\hbar n\omega)^{2}}{2},
\edq
and
\beq
f_\alpha(\veps)\sim f(\veps)-\partial_\veps f \delta\mu_\alpha+\partial^2_\veps f\frac{\delta\mu_\alpha^2}{2},
\edq
where $f(\veps)=[e^{(\veps-\mu)/(k_BT)}+1]^{-1}$ is the Fermi-Dirac distribution if we take the chemical potential $\mu$ and the base temperature $T$ as a reference.

In the slow driving regime, for which the typical frequency of the ac potential $\omega\rightarrow 0$, it is possible to do an exact analysis by expanding the Green function 
\beq\label{sl1}
G^{R}(t,\veps)=\sum_{n=-\infty}^{\infty}e^{-in\omega t}\calg(n,\veps),
\edq 
or equivalently the scattering matrix, in powers of $\omega$ \cite{us,mb,florlili}. 
By keeping terms up to first order in $\omega$ we get
\beq
\calg(n,\veps)\sim \calg^{(0)}(n,\veps)+\hbar\omega\calg^{(1)}(n,\veps).
\edq
{\color{red}}{
In the case of the driven single level, the above expression reduces to}
\beq\label{sl2}
G^{R}(t,\veps)=G^{f}(t,\veps)+\frac{i\hbar}{2}\partial_t\partial_\veps G^{f}(t,\veps)+...,
\edq
where $G^{f}=[\veps-\veps_d(t)+i\Gamma/2]^{-1}$ is the frozen Green function. 
Equation~(\ref{sl2}) is in fact quite general if the driving does not break the symmetry of scattering with respect to a spatial direction reversal~\cite{mbook}.

Then, combining Eqs. (\ref{sl1}) and (\ref{sl2}), we find
\ba
\calg^{(0)}(n,\veps)&=&\int_0^{\tau}\frac{dt}{\tau}G^{f}(t,\veps)e^{in\omega t}\nonumber\\
\omega\calg^{(1)}(n,\veps)&=&\int_0^{\tau}\frac{dt}{\tau}\frac{i}{2}\partial_t\partial_\veps G^{f}(t,\veps)e^{in\omega t}.
\ea

\section{Coefficients $\Lambda$ of the adiabatic expansion}\label{coefficients}
By using the low frequency expansion detailed in Appendix \ref{cq} in the expressions of the charge current and the power developed by the ac forces we can calculate
\ba\label{lambdas}
\Lambda^{C,\mu}&=&-\frac{1}{2}\int\frac{d\veps}{h}\partial_\veps f\sum_{\alpha=L,R} \Gamma_\alpha\rho^f_{\bar{\alpha}}\nonumber\\
\Lambda^{F,V}&=&\frac{\hbar}{2}\int\frac{d\veps}{2\pi}\partial_\veps f {\rho^f}^2,
\ea
where $f(\veps)=f_L(\veps)$, since we take the left reservoir as a reference.  We have used the notation
$\bar{L}=R$ and $\bar{R}=L$, as well as the property $\rho_\alpha^{f}=\vert G^{f}(t,\veps)\vert^{2}\Gamma_\alpha$ with $\alpha=L,R$. The local frozen density of states can be expressed as 
\beq
\rho^{f}(t,\veps)=-2\mbox{Im}\{G^{f}(t,\veps)\}=\vert G^{f}(t,\veps)\vert^{2}\Gamma,
\edq
 with $G^{f}(t,\veps)=[\veps-\veps_d(t)+i\Gamma/2]^{-1}$ being the frozen Green function describing the regime in which the electrons instantaneously adjust its potential to the ac field, and $\Gamma=\sum_{\alpha=L,R} \Gamma_\alpha$ is the total hybridization with the reservoirs.

Following a similar procedure in Eq. (\ref{heat}), we can compute $\dot{Q}_\alpha(t)$ up to second order in $\omega, \delta \mu$. The result is collected in the coefficients $\Lambda_\alpha$, which can be expressed as
\ba
& & \Lambda_\alpha^{\dot{V}}=-\int \frac{d\veps}{2\pi}\partial_\veps f (\veps-\mu)\rho_\alpha^f\nonumber\\
& & \Lambda_L^{\delta\mu}=-\Lambda_R^{\delta\mu}=\int \frac{d\veps}{h}\partial_\veps f (\veps-\mu)\Gamma_R\rho_L^f,
\ea
for the first order, and
\ba
& &  \Lambda_\alpha^{{\dot{V}}^2}=-\frac{\hbar}{2}\int \frac{d\veps}{2\pi}\partial_\veps f \partial_\veps[(\veps-\mu){\rho^f\rho^f_\alpha}] \nonumber\\
& & \Lambda_\alpha^{{\ddot{V}}}=\frac{\hbar}{2}\int \frac{d\veps}{2\pi}\partial_\veps f (\veps-\mu){\rho^f\rho^f_\alpha}\nonumber\\
& &  \Lambda_L^{\dot{V}\delta\mu}=\frac{1}{2}\int\frac{d\veps}{2\pi}\partial_\veps f \Gamma_R\partial_\veps [(\veps-\mu)\rho^f\rho^f_L]\\
& & \Lambda_R^{\dot{V}\delta\mu}=\int\frac{d\veps}{2\pi}\partial_\veps f(\veps-\mu) \partial_\veps \rho^f_R\,-\Lambda_L^{\dot{V}\delta\mu}\nonumber\\
& &  \Lambda_L^{\delta\mu^2}=-\frac{1}{2}\int\frac{d\veps}{h}\partial_\veps f \Gamma_R\partial_\veps[(\veps-\mu)\rho^f_L]\nonumber\\
& &  \Lambda_R^{\delta\mu^2}=- \Lambda_L^{\delta\mu^2} +\Lambda^{C,\mu},\nonumber
\ea
for the second order.

Similarly, the coefficients entering the expansion of $\dot{Q}_S(t)$ read
\ba
& &\Lambda_S^{\dot{V}}=-(\Lambda_L^{\dot{V}}+\Lambda_R^{\dot{V}})=\int \frac{d\veps}{2\pi}\partial_\veps f (\veps-\mu)\rho^f \nonumber\\
& &\Lambda_S^{{\dot{V}}^2}=-(\Lambda_L^{{\dot{V}}^2}+\Lambda_R^{{\dot{V}}^2}+\Lambda^{F,V})=\frac{\hbar}{2}\int \frac{d\veps}{2\pi}\partial_\veps f (\veps-\mu)\partial_\veps{\rho^f}^2\nonumber\\
& &\Lambda_S^{\ddot{V}}=-(\Lambda_L^{\ddot{V}}+\Lambda_R^{\ddot{V}})=-\frac{\hbar}{2}\int \frac{d\veps}{2\pi}\partial_\veps f (\veps-\mu){\rho^f}^2\\
& &\Lambda_S^{\dot{V}\delta\mu}=-(\Lambda_L^{\dot{V}\delta\mu}+\Lambda_R^{\dot{V}\delta\mu})=-\int \frac{d\veps}{2\pi}\partial_\veps f (\veps-\mu)\partial_\veps\rho_R^f.\nonumber
\ea


\begin{thebibliography}{99}

%
%
%
%
%
%
%
%
%
%
%
%
%
%
%
%
%
%
%
%
%
%
%
%
%
%
%
%
\bibitem{cold}J-P Brantut, C. Grenier, J.  Meineke, D. Stadler, S. Krinner, C. Kollath, T. Esslinger, and A. Georges,  Science {\bf 342}, 713 (2013)

\bibitem{nanomec1}G.A. Steele, A.K. H\"uttel, B. Witkamp, M. Poot, H.B. Meerwaldt, L.P. Kouwenhoven, and H.S.J. van der Zant, Science \textbf{325}, 1103 (2009).

\bibitem{nanomec2} C. Lotze, M. Corso, K. J. Franke, F. von Oppen, J. I. Pascual, Science {\bf 338}, 779 (2012)

\bibitem{Michelini:2016ju}
F. Michelini, K. Beltako, A. Cr\'epieux, SPIE Newsroom (2016), \doi{10.1117/2.1201601.006320}.


\bibitem{elec1}D. Segal and A. Nitzan, Phys. Rev. E \textbf{73}, 026109 (2006).

\bibitem{elec2}L. Arrachea, M. Moskalets, and L. Martin-Moreno, Phys. Rev. B \textbf{75}, 245420 (2007).

\bibitem{wu09}
L.-A. Wu and D. Segal, J. Phys. A: Math. Theor. \textbf{42}, 025302 (2009).

\bibitem{elec4} R. Bustos-Marun, G. Refael and F. von Oppen, Phys. Rev. Lett. {\bf 111} 060802 (2013)

\bibitem{elec3}R. Biele, R. D'Agosta and A. Rubio, Phys. Rev. Lett. {\bf 115}, 056801 (2015)

\bibitem{Carrega:2015dh}
M. Carrega, P. Solinas, A. Braggio, M. Sassetti, and U. Weiss,  New Journal of Physics {\bf 17}, 045030 (2015).

\bibitem{Chen:2015hu}
J. Chen, M. ShangGuan, and J. Wang, New Journal of Physics {\bf 17}, 053034 (2015).

\bibitem{camp:2015}
	M. Campisi, J. Pekola, and R. Fazio, New Journal of Physics {\bf 17}, 035012 (2015).

\bibitem{thermolud} M.F. Ludovico, F. Battista, F. von Oppen and L. Arrachea, Phys Rev B {\bf 93}, 075136 (2016).

\bibitem{Proesmans:2016tq}
K. Proesmans, B. Cleuren, and C. Van den Broeck, J. Stat. Mech. {\bf 2016}, 023202 (2016).

 \bibitem{Dare:2016jk}
 A. M. Dar\'e and P. Lombardo, Phys. Rev. B {\bf 93}, 035303 (2016).

\bibitem{Bruch:2015ux}
 A. Bruch, M. Thomas, S. Viola Kusminskiy, F. von Oppen, and A. Nitzan,
 Phys. Rev. B \textbf{93}, 115318 (2016).


\bibitem{jar} C. Jarzynski, Phys. Rev. Lett.
{\bf 78}, 2690 (1997).

\bibitem{crook}G. E. Crooks, Phys. Rev. E
{\bf 60}, 2721 (1999).

\bibitem{fluc1} J. Kurchan, J. Stat. Mech. P07005 (2007).

\bibitem{fluc2}K. Saito and A. Dhar, Phys. Rev. Lett. {\bf  99}, 180601 (2007).

\bibitem{sai08}
K. Saito and Y. Utsumi, Phys. Rev. B \textbf{78}, 115429 (2008).
\bibitem{for08}
H. F\"orster and M. B\"uttiker, Phys. Rev. Lett. \textbf{101}, 136805 (2008).
\bibitem{san09}
D. S\'anchez, Phys. Rev. B \textbf{79}, 045305 (2009).

\bibitem{fluc3}M.  Esposito,  U.  Harbola,  and  S.  Mukamel,  Rev.  Mod. Phys.
{\bf 81}, 1665 (2009).

\bibitem{fluc4}M. Esposito, R. Kawai, K. Lindenberg and C. Van den Broeck, Phys. Rev. Lett.  {\bf 105}, 150603 (2010). 

\bibitem{nak10}
S. Nakamura, Y. Yamauchi, M. Hashisaka, K. Chida, K. Kobayashi, T. Ono, R. Leturcq, K. Ensslin, K. Saito, Y. Utsumi, and A. C. Gossard, Phys. Rev. Lett. \textbf{104}, 080602 (2010).


\bibitem{fluc5}M. Campisi, P. H\"anggi, and P. Talkner, Rev. Mod. Phys.
{\bf 83}, 771 (2011).

\bibitem{lop12}
R. L\'opez, J. S. Lim, and D. S\'anchez,
Phys. Rev. Lett. \textbf{108}, 246603 (2012).

\bibitem{wan13}
C. Wang and D. E. Feldman,
Phys. Rev. Lett. \textbf{110}, 030602 (2013)

\bibitem{uts14}
Y. Utsumi, O. Entin-Wohlman, A. Aharony, T. Kubo, and Y. Tokura,
Phys. Rev. B \textbf{89}, 205314 (2014).

\bibitem{thermo1}R. Uzdin, A. Levy, and R. Kosloff, Phys. Rev. X
{\bf 5}, 031044 (2015).

\bibitem{thermo2}K. Brandner and U. Seifert, ArXiv:1604.03411 (2016).

\bibitem{thermo3} U. Seifert,  Phys. Rev. Lett. {\bf 116}, 020601 (2016).


\bibitem{heat1}D. Segal, Phys. Rev E {\bf 90}, 012148 (2014). 

\bibitem{heat2}M. Carrega, P. Solinas, M. Sassetti, and U. Weiss, Phys. Rev. Lett. \textbf{116}, 240403 (2016). 

\bibitem{heat3}R.  Schmidt,  M.  F.  Carusela,  J.  P.  Pekola,  S.  Suomela,    and
J. Ankerhold, Phys. Rev. B
{\bf 91}, 224303 (2015). 

\bibitem{heat4}L. Arrachea, E. R. Mucciolo, C. Chamon,   and R. B. Capaz, Phys. Rev. B
{\bf 86}, 125424 (2012).

\bibitem{heat5}R. Tuovinen, N. S\" akkinen, D. Karlsson, G. Stefanucci and R. van Leeuwen, Phys. Rev. B {\bf 93}, 214301 (2016).

\bibitem{heat6}N. Beraha, A. Soba and  M. F. Carusela,  arXiv:1604.07714 (2016).




\bibitem{us} M. F. Ludovico, J. S. Lim, M. Moskalets, L. Arrachea, and D. S\'anchez,
Phys. Rev. B  \textbf{89}, 161306(R) (2014).

\bibitem{balian}
R. Balian, \textit{From Microphysics to Macrophysics} (Springer-Verlag, Berlin, 1982), vol. I, ch. 5.

\bibitem{ramer}J. Rammer, \textit{Quantum Field Theory of Non-equilibrium States
} (Cambridge University Press, New York, 2007).
\bibitem{defq}
G. D. Mahan, \textit{Many-Particle Physics} (Kluwer/Plenum, New York, 2000), ch. 3.

\bibitem{Ankerhold:2014ky}
J. Ankerhold and J. P. Pekola, Phys. Rev. B {\bf 90}, 075421 (2014).

\bibitem{Esposito:2015bg}
M. Esposito, M. A. Ochoa, and M. Galperin, Phys. Rev. Lett. {\bf 114}, 080602 (2015).






\bibitem{guillem}
G. Rossell\'o, R. L\'opez, and J. S. Lim,
Phys. Rev. B \textbf{92}, 115402 (2015).

\bibitem{lili1} L. Arrachea, Phys. Rev. B {\bf 72},  125349 (2005)

\bibitem{lili2} L. Arrachea,
	Phys. Rev. B, vol. {\bf 75},  035319 (2007)

\bibitem{lilimoskalets} L. Arrachea and M. Moskalets, Phys. Rev. B {\bf 74}, 245322 (2006).

\bibitem{Buttiker:1993bw}
M. B\"{u}ttiker, H. Thomas, A. Pr\^{e}tre, Mesoscopic capacitors, Phys. Lett. A {\bf 180}, 364--369 (1993).


\bibitem{Gabelli:2006eg}
J. Gabelli, G. F\`eve, J.-M. Berroir, B. Pla\c{c}ais, A. Cavanna, E. al, Y. Jin, and D. C. Glattli, Science {\bf 313}, 499--502 (2006).




\bibitem{Nigg:2006kl}
S. E. Nigg, R.  L\'opez, and M. B{\"u}ttiker, Phys. Rev. Lett. {\bf 97}, 206804 (2006).

\bibitem{Mora:2010hw}
C. Mora and K. Le Hur, Nature Phys. {\bf 6}, 697 (2010).

\bibitem{Filippone:2011fy}
M. Filippone, K. Le Hur, and C. Mora, Phys. Rev. Lett. {\bf 107}, 176601 (2011).

%
%
%
%
%
%
%
%
%
%
%
%
%
%
%

\bibitem{Buttiker:1992ge}
M. B\"{u}ttiker, Phys. Rev. B {\bf 46}, 12485 (1992).


\bibitem{mbook} M. Moskalets, \textit{Scattering matrix Approach to Non-Stationary Quantum Transport}, World Scientific, London, 2012.

\bibitem{Battista:2013ew}
F. Battista, M. Moskalets, M. Albert, and P. Samuelsson, Phys. Rev. Lett. {\bf 110}, 126602 (2013).

\bibitem{Lim:2013de}
 J. S. Lim, R. L\'opez, and D. S\'anchez, Phys. Rev. B {\bf 88}, 201304 (2013).

\bibitem{Battista:2014di}
F. Battista, F. Haupt, and J. Splettstoesser, Phys. Rev. B {\bf 90}, 085418 (2014).



\bibitem{Fisher-Lee} D.S. Fisher and P.A. Lee, Phys. Rev. B {\bf 23}, 6851 (1981).

\bibitem{mb} M. Moskalets and M. B{\"u}ttiker, Phys. Rev. B
{\bf 69}, 205316 (2004).

\bibitem{florlili}M. F. Ludovico and L. Arrachea, Phys. Rev. B {\bf 87}, 115408
(2013).

\end{thebibliography}
\end{document}